\documentclass[showpacs,amsmath,amssymb,pra,superscriptaddress,floatfix,twocolumn]{revtex4-1}

\usepackage{longtable}
\usepackage{amssymb, graphicx,subfigure}
\usepackage{enumerate,tensor}
\usepackage{enumerate}
\usepackage{nicefrac}
\usepackage{amsmath,bm}
\usepackage{hyperref}
\usepackage{epsfig}
 \usepackage{relsize}

\newcommand{\ie}{i.\,e.}%
\def\pra#1#2#3{Phys.~Rev.~A~{\bf #1},\ #2\ (#3)}

\def\bea{\begin{eqnarray}} 
\def\eea{\end{eqnarray}} 
\def\be{\begin{equation}} 
\def\ee{\end{equation}}

\makeatletter
\def\subsubsection{\@startsection{subsubsection}{3}{10pt}{-1.25ex plus -1ex minus -.1ex}{0ex plus 0ex}{\normalsize\bf}}
\def\paragraph{\@startsection{paragraph}{4}{10pt}{-1.25ex plus -1ex minus -.1ex}{0ex plus 0ex}{\normalsize\textit}}
\renewcommand\@biblabel[1]{#1}
\renewcommand\@makefntext[1]%
{\noindent\makebox[0pt][r]{\@thefnmark\,}#1}
\DeclareRobustCommand\onlinecite{\@onlinecite}
\def\@onlinecite#1{\begingroup\let\@cite\NAT@citenum\citealp{#1}\endgroup}
\def\tagform@#1{\maketag@@@{\ignorespaces#1\unskip\@@italiccorr}}
\let\orgtheequation\theequation
\def\theequation{(\orgtheequation)}
\makeatother

\newcommand{\ry}{Rydberg }

\begin{document}

\title{Rotational hybridization, and control of alignment and orientation in triatomic ultralong-range Rydberg molecules}

\author{Rosario Gonz\'alez-F\'erez}
\email{rogonzal@ugr.es}
\affiliation{Instituto Carlos I de F\'{\i}sica Te\'orica y Computacional,
and Departamento de F\'{\i}sica At\'omica, Molecular y Nuclear,
  Universidad de Granada, 18071 Granada, Spain}
\affiliation{The Hamburg Center for Ultrafast Imaging, Luruper Chaussee 149, 22761 Hamburg, Germany}
\author{H. R. Sadeghpour}

\affiliation{ITAMP, Harvard-Smithsonian Center for Astrophysics, Cambridge, Massachusetts 02138, USA} 
  
\author{Peter Schmelcher}
\affiliation{The Hamburg Center for Ultrafast Imaging, Luruper Chaussee 149, 22761 Hamburg, Germany}
\affiliation{Zentrum f\"ur Optische Quantentechnologien, Universit\"at
  Hamburg, Germany} 
\date{\today}

\begin{abstract}
We explore the electronic structure and rovibrational properties of an ultralong-range triatomic \ry  molecule
formed by a \ry atom and a ground state heteronuclear diatomic molecule. We focus here on interaction of Rb($27s$) \ry atom with KRb($N=0$) diatomic polar molecule. There's significant electronic hybridization of Rb($n=24$, $l\ge 3$) degenerate manifold. 
The polar diatomic molecule is allowed to rotate in the electric fields generated by the
Rydberg electron and core as well as an external field. We investigate the metamorphosis of the
Born-Oppenheimer potential curves, essential for the binding of the molecule, with varying electric field
and analyze the resulting properties such as the vibrational structure and the alignment and orientation of the polar diatomic molecule. 
\end{abstract}

\maketitle
\section{Introduction}
\label{sec:intro}

While in the early years of Rydberg physics there was a substantial focus on high resolution spectroscopy \cite{stebbings83}
the advent of Bose-Einstein condensation \cite{pethick08} has brought Rydberg atoms and molecules into the focus of fields
that were previously not considering these highly excited and fragile quantum objects. This includes Rydberg gases and
plasmas \cite{pohl11}, quantum optics and many-body physics of long-range interacting Rydberg systems \cite{pohl11}, quantum simulators
\cite{weimer10} as well as quantum information processing based on entanglement generation between interacting Rydberg atoms in optical lattices
\cite{saffman10}. A very recent spectacular development is the prediction \cite{greene00} and experimental detection
\cite{bendkowsky09} of ultralong-range molecules from an ultracold atomic cloud. 
Opposite to the well-known Rydberg molecules with a tightly bound positively charged core and an electron circulating around
it, the ultralong-range molecular species follows a very much different bonding mechanism with, for a diatomic molecule, internuclear
distances of the order of the extension of the electronic Rydberg state. Here, 
a ground state atom locally probes the Rydberg electronic wavefunction whose interaction is typically described by s- and p-wave
Fermi-type pseudopotentials \cite{fermi34,omont77}. The latter leads to the unusual oscillatory behaviour of the corresponding adiabatic
potential energy curves and a correspondingly rich vibrational dynamics. For principal quantum numbers of typically $n \approx 30-100$,
being routinely prepared in the experiments, the diatomic molecules can be up to a few $\mu m$ in size. These exotic species were first
predicted theoretically back in 2000 by Greene {\it{et al}} \cite{greene00} providing polar (trilobite) and non-polar states 
with vibrational binding energies in the GHz and MHz regime, correspondingly.  
Beyond the detection of non-polar ultralong-range molecules and their spectroscopic characterization
Rydberg triatomic molecules and excited diatomic molecules bound by quantum reflection have been found \cite{bendkowsky10}.

The sensitivity of Rydberg atoms to external fields carries over also to ultralong-range Rydberg molecules which opens unprecedented
opportunities for their weak field control. This includes static magnetic and electric fields or laser fields
which could be used to change the electronic structure severely and correspondingly change the molecular geometry
and rovibrational dynamics. Electric and magnetic field control of molecular binding properties \cite{lesanovsky07,kurz12,kurz13}
as well as alignment have been demonstrated very recently \cite{krupp14}.

If the neutral ground state atom immersed into the Rydberg wave function by a
$\Lambda$-doublet  heteronuclear diatomic molecule~\cite{rittenhouse10,rittenhouse11,mayle12}, giant polyatomic \ry molecules can form. These  triatomic ultralong-range \ry molecules
could be created in an ultracold mixture of atoms and molecules by using standard \ry excitation schemes.  
The \ry electron is coupled to the two internal states of the polar ground state molecule, creating a series of undulating Born-Oppenheimer potential curves (BOP) due to the oscillating character of the \ry electron wave functions. 
The electronic structure of these giant molecules could be easily manipulated by means of electric fields of a few V/cm,
creating a complex set of avoided crossings among neighboring levels. In particular   
a Raman scheme to couple the two internal states of opposite orientation of the polar diatomic molecule~\cite{rittenhouse10}
has been proposed.

In the present work, we consider a triatomic molecule formed by a Rb \ry atom and a ground state KRb polar molecule.
Compared to the previous works~\cite{rittenhouse10,rittenhouse11,mayle12}, 
we perform an extended and realistic treatment of the internal motion of the diatomic molecule. 
Indeed, we explicitly introduce the angular degrees of freedom of the diatomic molecule and as such are capable of
describing properly the molecular rotational in the rigid rotor approximation. 
 The free rotation of the diatomic molecule in the triatomic \ry molecule is now dressed due to the combined electric fields of  the \ry electron and the core, and is characterized by a strong orientation and alignment along the field direction. 
Since the permanent dipole moment of the molecule is not fixed in space but rotates, we include in
our description the three components of the  electric field induced by the Rydberg atom.
The perpendicular components of the electric field to the space-fixed frame along the $Z$ axis ensure that the
magnetic quantum numbers of the electron and the polar diatomic molecule to no longer be conserved. 
We analyze the BOP of the Rb(ns)-KRb \ry triatomic molecule as the internuclear separation between 
KRb and  the Rb$^+$ core varies and explore in detail the effects of the electric field induced by the \ry
atom on the rotational motion of KRb, by studying the hybridization of its angular motion, orientation and alignment. 
We also investigate the impact of an additional electric field on these BOP and their vibrational spectrum.  

The paper is organized as follows: In \autoref{sec:hamil} we describe the Hamiltonian of the system 
and the electric field induced by the \ry core and electron in the diatomic molecule.
The BOP as a function of the distance between the perturbing diatomic molecule and the \ry core as well as the resulting
alignment and orientation of the diatomic molecule are presented in   
\autoref{sec:ffBO}.  In  \autoref{sec:efd_BO_p}, we focus on the impact of an external field.
The vibrational spectrum of the \ry triatomic molecule is discussed in detail in  \autoref{sec:BO_p_vib} and
the conclusions are provided in \autoref{sec:con}.

\section{Hamiltonian, interactions and computational approach}
\label{sec:hamil}

We consider a  triatomic molecule formed by  a Rydberg atom and a ground state heteronuclear diatomic molecule
in an electric field. 
The permanent electric dipole moment of the polar molecule interacts with the electric field provided
by the \ry core and electron as well as the external electric field.
With increasing electric field strength the rotational motion becomes a liberating one which
leads to the alignment and orientation of the molecular axis.
Here, we describe the diatomic molecule by means of a rigid rotor approach. This 
should be a good approximation for the tightly bound molecule since we are allowed to adiabatically separate 
the rotational and vibrational motions even in the presence of the electric field induced
by the  \ry atom~\cite{gonzalez04,gonzalez05}. 

\begin{figure}[h]
\begin{center}
\includegraphics[angle=0,width=0.8\columnwidth]{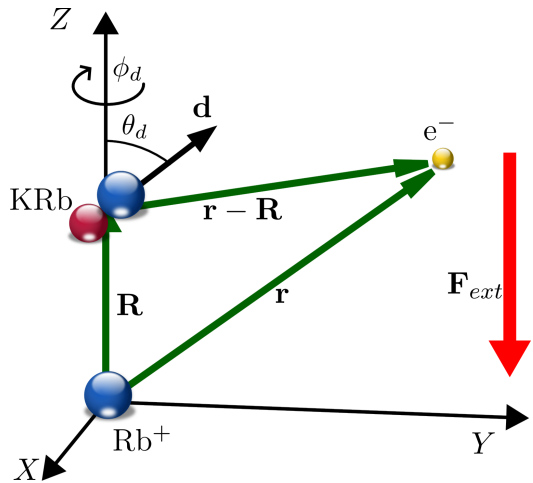}
\caption{
A  sketch (not to scale) of the triatomic molecule formed by the Rydberg atom
and diatomic polar molecule is shown.}%
\label{fig:molecule}
\end{center}
\end{figure}

For the sake of simplicity, we fix the geometry of the \ry triatomic molecule: 
The laboratory fixed frame (LFF)  is defined so that the Rydberg core is located in its origin, and 
the  diatomic molecule is the $Z$-axis at a distance $R$. A qualitative sketch is presented in \autoref{fig:molecule}
The positions of the diatomic molecule and electron in the LFF 
are $\mathbf{R}=R\hat{Z}$ and $\mathbf{r}=(r\hat{R},\theta_r\hat{\theta}, \phi_r \hat{\phi})$, respectively.

Within the Born-Oppenheimer approximation, the adiabatic Hamiltonian of this system is given by
\begin{equation}
\label{eq:Hamil_adiabatic}
H_{ad}=H_A+H_{mol} + H_{int}.  
\end{equation}
The first term stands for the single electron Hamiltonian describing the \ry atom 
\begin{equation}
\label{eq:Hamil_atom}
H_A=-\frac{\hbar^2}{2m_e}\nabla^2_{r}+V_l(r)
\end{equation}
with $V_l(r)$ being the  $l$-dependent model potential where $l$ is the angular momentum quantum number
of the \ry electron \cite{msd94}.%

The second term $H_{mol}$ is the Hamiltonian of the polar molecule in the electric field 
created by the Rydberg electron and core, $\mathbf{F}_{ryd}(\mathbf{R},\mathbf{r})$.
In the rigid rotor approximation, the molecular Hamiltonian reads 
\begin{equation}
\label{eq:Hamil_molecule}
H_{mol}=B\mathbf{N}^2-\mathbf{d}\cdot\mathbf{F}_{ryd}(\mathbf{R},\mathbf{r}) 
\end{equation}
with $B$ being the rotational constant, $\mathbf{N}$ the molecular angular momentum operator 
and $\mathbf{d}$ the permanent electric dipole moment of the diatomic molecule, which is 
parallel to the molecular internuclear axis.
Note that the internal rotational motion of the diatomic molecule is described by the Euler angles 
$\Omega_d=(\theta_d,\phi_d)$,
which relate the diatomic molecular fixed frame (MFF) and LFF. 
The MFF is defined with its origin at the center of mass of the two nuclei of this diatomic molecule and 
the $Z_M$ axis being along the internuclear axis. The electric field due to the Rydberg electron and core is given by 
\begin{equation}
\label{eq:field_rydberg_e_core}
\mathbf{F}_{ryd}(\mathbf{R},\mathbf{r})=e\frac{\mathbf{R}}{R^3}+e\frac{\mathbf{r}-\mathbf{R}}{|\mathbf{r}-\mathbf{R}|^3}
\end{equation}
where $e$ is the electron charge \cite{rittenhouse10,rittenhouse11,mayle12}.

The last term in the adiabatic Hamiltonian \eqref{eq:Hamil_adiabatic} stands for the interaction of the \ry atom and the diatomic molecule 
 with the external electric field $\mathbf{F}_{ext}$
\begin{equation}
\label{eq:Hamil_ex_field_inter}
 H_{int}=e\mathbf{F}_{ext}\cdot\mathbf{r}- \mathbf{d}\cdot\mathbf{F}_{ext}.
\end{equation}
Here, we consider an external electric field antiparallel along the LFF $Z$-axis  with strength
 ${F}_{ext}$, \ie, $\mathbf{F}_{ext}= -F_{ext} \hat{Z}$. 

 The total angular momentum of the triatomic molecule, but excluding an overall rotation, is given by $\mathbf{J}=\mathbf{l}+\mathbf{N}$,
where $\mathbf{l}$ is the orbital angular momentum of the \ry electron. 
In the absence of the external electric field,
the triatomic molecule states can be characterized  by the projection of  $\mathbf{J}$ along the LFF Z-axis, \ie,  $M_J$. 
In the presence of the external field $\mathbf{F}_{ext}=-{F}_{ext}\hat{Z}$, there's 
azimuthal symmetry and $M_J$
is conserved. In both cases, for $M_J\ne0$, states with $M_J$ and $-M_J$ are degenerate.

To solve
the Schr\"odinger equation associated with the Hamiltonian \eqref{eq:Hamil_adiabatic}, we perform a basis set expansion 
in terms of the coupled basis 
\begin{eqnarray}
%\label{eq:coupled_basis}
 | nlNJM_J\rangle&=&
\Psi_{nlNJM_J}(\mathbf{r},\Omega_d)=
\sum_{m_l=-l}^{m_l=l}\sum_{M_N=-N}^{M_N=N}\psi_{nlm}(\mathbf{r}) \nonumber\\
&& \times Y_{NM_N}(\Omega_d)
\langle l m_l NM_N| J M_J\rangle
\label{eq:coupled_basis}
\end{eqnarray}
where $\langle l m_l N M_N| J M_N\rangle$ is the Clebsch-Gordan coefficient,
$J=|l-N|,\dots,l+N$,  and  $M_J=-J,\dots,J$. 
$\psi_{nlm}(\mathbf{r})$ is the \ry electron wave function
with $n$, $l$ and $m$ being the principal, orbital and magnetic quantum numbers, respectively.
$Y_{NM_N}(\Omega_d)$ is the   field-free rotational wave function of the diatomic molecule,
with  $N$ and $M_N$ being the rotational and magnetic quantum numbers,
\ie, $Y_{NM_N}(\Omega_d)$ are the spherical harmonics. 
%Due to the symmetries of the \ry triatomic molecule, we solve the Sch\"odinger equation for fixed value of the magnetic quantum number $M_{J}$. 

In the computation of the Hamiltonian matrix $\langle nlNJM_J |H_{mol}|n'l'N' J'M_{J}\rangle$
we determine the matrix elements $\langle nlm|\mathbf{F}_{ryd}(\mathbf{R},\mathbf{r})|n'l'm'\rangle$ using 
the relation
\begin{equation}
\label{eq:gradiente}
\frac{\mathbf{r}-\mathbf{R}}{|\mathbf{r}-\mathbf{R}|^3}=\nabla_{\mathbf{R}}\frac{1}{|\mathbf{r}-\mathbf{R}|}
\end{equation}
where $\mathbf{\nabla}_{\mathbf{R}}$ is the Laplacian with respect to the molecular 
coordinate
$\mathbf{R}=(R\hat{R},\theta_R\hat{\theta}, \phi_R \hat{\phi})$, see Ref. \cite{ayuel09}. 
The electric field reads
\begin{equation}
\mathbf{F}_{ryd}(\mathbf{R},\mathbf{r})=\frac{e\hat{Z}}{R^2}+e\sum_{l=0}^\infty\sqrt{\frac{4\pi}{2l+1}}\mathbf{A}_l(R,r,\Omega)
\end{equation}
with
\begin{widetext}
\begin{equation}
\label{eq:expre_A}
\mathbf{A}_l(R,r,\Omega)=\left\{
\begin{array}{l}
\frac{r^l}{R^{l+2}}\Big[
\frac{\sqrt{l(l+1)}}{2}Y_{l1}^-(\Omega)\hat{X}
-\frac{i\sqrt{l(l+1)}}{2}Y_{l1}^+(\Omega)\hat{Y}-(l+1)Y_{l0}(\Omega)\hat{Z}\Big]   \quad
\text{if}\quad r<R, 
\\
\frac{R^{l-1}}{r^{l+1}}\Big[
\frac{\sqrt{l(l+1)}}{2}Y_{l1}^-(\Omega)\hat{X}
-\frac{i\sqrt{l(l+1)}}{2}Y_{l1}^+(\Omega)\hat{Y}+
lY_{l0}(\Omega)\hat{Z}\Big]   \quad \text{if}\quad r>R,
\end{array}\right.
\end{equation}
\end{widetext}
where 
$\Omega=(\theta, \phi )$ and $\mathbf{r}=(r\hat{R},\theta\hat{\theta}, \phi \hat{\phi})$ are the coordinates of 
the \ry electron, and
\begin{equation}
Y_{l1}^\pm(\Omega)=\Big(Y_{l1}(\Omega)\pm Y_{l-1}(\Omega)\Big).
\end{equation}
In \autoref{eq:expre_A}, $\mathbf{R}=R\mathbf{Z}$. The first two terms in  expression \eqref{eq:expre_A} couple states with $\Delta m_l=\pm 1$, which means that
the projection $m_l$ on the $Z$ axis is not conserved. 
In addition, the  electric field induced by the \ry electron and the core  is non-parallel to the $Z$-axis, implying that 
for the diatomic molecule $M_N$ is  not  a good quantum number.
The matrix elements of  $\mathbf{F}_{ryd}(\mathbf{R},\mathbf{r})$ in the \ry electron basis are given by 
\begin{widetext}
\begin{equation}
\langle nlm|\mathbf{F}_{ryd}(\mathbf{R},\mathbf{r})|n'l'm'\rangle=
F_{nlmn'l'm'}^X(R)\hat{X}+
F_{nlmn'l'm'}^Y(R)\hat{Y}+
F_{nlmn'l'm'}^Z(R)\hat{Z}
\end{equation}
with 
\begin{equation}
F_{nlmn'l'm'}^X(R)=
e \sqrt{2\pi}\sum_{l''=|l-l'|}^{l+l'}
%\sqrt{\frac{4\pi}{2l''+1}}\frac{\sqrt{l''(l''+1)}}{2}
\sqrt{2\pi}\sqrt{\frac{l''(l''+1)}{2l''+1}}
{\cal{R}}_{l'',nln'l'}(R)
\int_{\Omega} Y_{l'm'}^*(\Omega)Y_{l''1}^-(\Omega)Y_{lm}(\Omega)d\Omega,
\label{eq:fx}
\end{equation}
\begin{equation}
F_{nlmn'l'm'}^Y(R)=
-ie \sqrt{2\pi}\sum_{l''=|l-l'|}^{l+l'}
%\sqrt{\frac{4\pi}{2l''+1}}\frac{\sqrt{l''(l''+1)}}{2}
\sqrt{\frac{l''(l''+1)}{2l''+1}}
{\cal{R}}_{l'',nln'l'}(R)
\int_{\Omega} Y_{l'm'}^*(\Omega)Y_{l''1}^+(\Omega)Y_{lm}(\Omega)d\Omega
\label{eq:fy}
\end{equation}
and
\begin{equation}
F_{nlmn'l'm'}^Z(R)=\frac{e}{R^2}\delta_{nn'}\delta_{ll'}\delta_{mm'}+
e\sum_{l''=|l-l'|}^{l+l'}\sqrt{\frac{4\pi}{2l''+1}}{\cal{Z}}_{l'',nln'l'}(R)
\int_{\Omega} Y_{l'm'}^*(\Omega)Y_{l''0}(\Omega)Y_{lm}(\Omega)d\Omega. 
\label{eq:fz}
\end{equation}
The integral in Eq. \ref{eq:fz} ensures that $F_{nlmn'l'm'}^Z(R)$ is nonzero only if   
$|l-l'|\le l'' \le l+l'$.
The radial integrals take on the following appearance:
\begin{equation}
{\cal{R}}_{l'',nln'l'}(R)=
\int_0^R \frac{r^{l''}}{R^{l''+2}} \psi_{n'l'}^*(r)\psi_{nl}(r)r^2dr+
\int_R^\infty\frac{R^{l''-1}}{r^{l''+1}}\psi_{n'l'}^*(r)\psi_{nl}(r)r^2dr
\end{equation}
\begin{equation}
{\cal{Z}}_{l'',nln'l'}(R)=
-(l''+1)\int_0^R \frac{r^{l''}}{R^{l''+2}} \psi_{n'l'}^*(r)\psi_{nl}(r)r^2dr+
l''\int_R^\infty\frac{R^{l''-1}}{r^{l''+1}}\psi_{n'l'}^*(r)\psi_{nl}(r)r^2dr
\end{equation}
\end{widetext}
where $\psi_{nl}(r)$ is the radial component of the \ry electronic wave function. 

In the next sections, we consider a  \ry triatomic molecule formed by a  rubidium atom and the diatomic
molecule KRb. The rotational constant of  KRb  is $B= 1.114$~GHz ~\cite{ni09},
 and its electric dipole moment $d=0.566$~D~\cite{ni08}.
We solve the Schr\"odinger equation belonging to the Hamiltonian \eqref{eq:Hamil_adiabatic} 
using the coupled basis \eqref{eq:coupled_basis}, which includes Rb(n=24, l$\ge$ 3) degenerate manifold  and 
the energetically closest  neighboring Rydberg state $27s$. Note that we are neglecting the
quantum defect of the 24f \ry state.
For the diatomic molecule, we take into account are the rotational excitations for $N\le 6$.

 %%%%%%%%%%%%%%%%%%%%%%%
\section{The field-free Born-Oppenheimer potentials and properties}
\label{sec:ffBO}

In this section we investigate the BOP of the Rb-KRb triatomic molecule as a function of the separation $R$ 
between KRb and Rb$^+$. The electric field 
induced by the Rydberg core and electron on the diatomic molecule decreases towards zero 
as $R$ is increased.
When the polar diatomic molecule is located far enough from the  \ry ionic core and electron,
the system could be considered as formed by two subsystems: a \ry atom and the diatomic molecule. 
Thus, for $R>>1~a_0$,  the BOP of the triatomic molecule approach the field-free limit
$E_{nl}+BN(N+1)$, with $N$ being the field-free rotational quantum number of the diatomic molecule  and 
$E_{n,l}$ the energy of the \ry atom in a state with principal and orbital quantum numbers $n$ and $l$.

\begin{figure}[h]
\begin{center}
\includegraphics[angle=0,width=0.95\columnwidth]{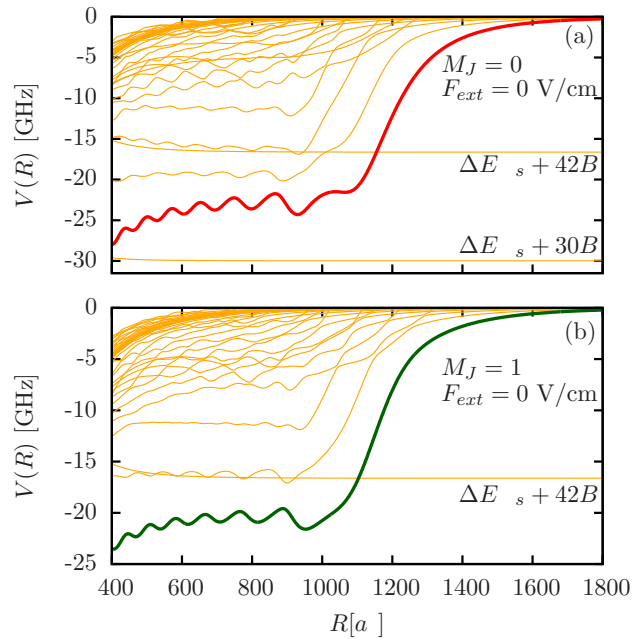}
\caption{Rb-KRb triatomic \ry molecule: 
Born-Oppenheimer potentials as a function of the separation between the
Rydberg core and the polar molecule $R$ for (a) $M_J=0$  and (b) $M_J=1$.
The calculations include the $27$s state and the degenerate manifold
$n=24$ and $l\ge3$ of rubidium. 
The lowest lying states with $M_J=0$ and $1$ evolving from the $n=24$ and $l\ge3$ manifold are plotted with thicker red and green
lines, respectively.
%On the right hand side of both panels, we have indicated the field-free 
%energy of the two asymptotes.
}%
\label{fig:BOP_MF_0_1_F0}
\end{center}
\end{figure}

In \autoref{fig:BOP_MF_0_1_F0}~(a) and \autoref{fig:BOP_MF_0_1_F0}~(b), 
we present the BOP evolving from the Rb($n=24$, $l\ge 3$) manifold 
with $M_J=0$ and $M_J=1$, respectively, as a function of $R$. 
Here, the zero energy has been set to the energy of Rb($n=24$, $l\ge 3$) degenerate manifold and KRb($N=0$) level. All the curves within a panel belong to triatomic molecule states with the 
same symmetry, and all the crossings between the BOP are therefore avoided crossings.

Two different behaviors are observed in \autoref{fig:BOP_MF_0_1_F0}. The BOP of the triatomic molecule levels formed from the Rb($27s$) state 
and KRb in an excited state  show a  weak dependence on $R$.  
Their field-free limits are $\Delta E_{27s}+30B$ %=|E_{27s}-E_{24,3}|+30B$
and  $\Delta E_{27s}+42B$,
with $\Delta E_{27s}=|E_{27s}-E_{24,3}|$,  referring to  KRb($N=5$, $6$ )rotational excitations, respectively.  
In contrast, the BOP of the triatomic molecule levels formed from the Rb($n=24$, $l\ge3$) manifold
have an oscillatory behaviour as a function of $R$. 
This reflects the dominant oscillatory behavior of the \ry electric field 
which is due to the  oscillatory structure  of the \ry electron wave functions.

The lowest triatomic molecule states with $M_J=0$ and $M_J=1$
steming from Rb($n=24$, $l\ge3$) \ry manifold are shifted more than $20$~GHz for $R\lesssim 1200~a_0$ and
$1000~a_0$, respectively.
When $R$ is increased beyond a certain limit,  their energies increase and approach the
field-free dissociation limit, Rb($n=24$, $l\ge3$) - KRb($N=0$).  
%\ry manifold and  KRb in its ground state $N=0$. 

These results are converged with respect to the size of the basis set in  \eqref{eq:coupled_basis}.  
\autoref{tb:energies_Rb} contains the energies of the \ry levels close to the Rb($n=24$, 
$\l\ge 3$) degenerate manifold. The energy gaps between the \ry levels 
are larger than the energy shifts of the levels evolving from this degenerate 
manifold observed in Rb-KRb, see  \autoref{fig:BOP_MF_0_1_F0}.  
This justifies that only the next neighboring level 27s has to be included in the basis set.
Further including the 25d and 26p states decreases the energy differences to less than $1\%$.
It may be possible for Rb-KRb triatomic molecule levels 
 formed from the KRb diatomic molecule in highly excited rotational states and a Rb atom 
 in lower lying \ry levels to produce BOP within the spectral range of \autoref{fig:BOP_MF_0_1_F0}. 
However, the coupling of
such triatomic molecule states to those presented  in \autoref{fig:BOP_MF_0_1_F0} would be weak, because of the large rotational energy separation in KRb. 

\begin{table}[htdp]
\caption{
Atomic rubidium: energies and energy differences  $ \Delta E_{nl}=|E_{n,l}-E_{24,3}|$ 
of the \ry levels close to the degenerate Rydberg manifold $n=24$ and $l\ge3$. 
\label{tb:energies_Rb}}
\begin{center}
\begin{tabular}{cc|c|c}
 \hline
% n-manifold above n=          24
$n$ & $l$ & $E_{n,l} $ [THz] & $ \Delta E_{nl}$[GHz] \\ 
%|E_{n,l}-E_{24,3}|$  [GHz] \\ 
 \hline
$25 $ & $3$ & $-5.26375$ & $   447.78$ \\
$28 $ & $0$ & $-5.31981$ & $   391.72$ \\
$26 $ & $2$ & $-5.41266$ & $   298.87$ \\
$27 $ & $1$ & $-5.54796$ & $   163.57$ \\
 % n-manifold  n=          24
 \hline
$24 $ & $3$ & $-5.71153$ & $   0$ \\
$27 $ & $0$ & $-5.77493$ & $  63.40$ \\
$25 $ & $2$ & $-5.87995$ & $  168.42$ \\
$26 $ & $1$ & $-6.03333$ & $  321.80$ \\
  \hline
% n-manifold bellow n=          24 
%$23 $ & $3 $& $-6.21898$ & $  507.45$ \\
 \hline
\end{tabular}
\end{center}
\label{default}
\end{table}%

We have also checked the convergence of our results with the number of rotational 
states of KRb included in the basis set.  By increasing the number of rotational  excitations  from six to eight,  
the relative difference between the two sets of energies  
is smaller than  $2\times 10^{-6}$. This implies that BOP obtained with the rotational basis
including rotational excitations with $N\le 6$ 
are well converged. 

\begin{figure}[h]
\begin{center}
\includegraphics[angle=0,width=0.9\columnwidth]{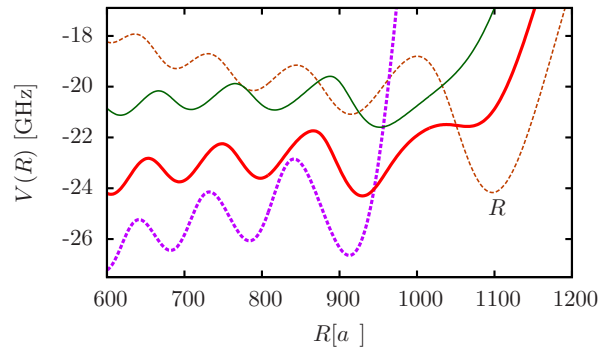}
\caption{For the Rb-KRb triatomic molecule,
Born-Oppenheimer potentials derived using the two state model,
states  L (thick dashed line) R (thin dashed line), and the rigid rotor description, levels 
with $M_J=0$ (thick solid line) and $M_J=1$ (thin solid line),  of the diatomic molecule KRb.
}%
\label{fig:BO_comparison}
\end{center}
\end{figure}
For comparison, we have also determined the BOP for a Rb-KRb triatomic molecule, but using a two-state approximation to describe the
internal motion of the diatomic molecule KRb~\cite{mayle12}.
In the two-state model, the rotational energy gap is obtained from the rotational constant,  $B=1.114$~GHz,
and the electric dipole moment,  
is parallel to the LFF $Z$-axis so that KRb cannot freely rotate. Thus, 
we only consider the $Z$ component of the electric
field induced by the \ry atom  $	\mathbf{F}_{ryd}$ in   \autoref{eq:fz}.
In \autoref{fig:BO_comparison}, we present the energies of the two lowest states
evolving from the Rb($n=24$, $l\ge 3$) manifold using the two-state model 
and the rigid rotor description. 
The energy shift  of these BOP from the Rb($n=24$, $l\ge3$) is smaller when 
the rigid rotor description is used. This could be due to the effect of the $X$ and $Y$ components of the electric field,  which 
reduce the net effect of the $Z$ component. 
The depth of the potential wells is also decreased, which implies that they will accommodate less vibrational levels.
This effect is particularly pronounced for the lowest wells of the two-state triatomic molecule near $R\sim 920~a_0$, and $1100~a_0$ 
 where the dipole is pointing toward and  away from the ion core, respectively. 
When the rigid-rotor approximation is used, the last potential well near $R\sim 1100~a_0$ is very shallow, and
the  diatomic molecule is also oriented away from Rb$^+$. In the two-state model these two  BOP
have different symmetries and  do cross, while the closest energy separation for the rigid rotor description is approximately $1.8$~GHz.
Furthermore, they are not coupled  if an additional electric field is applied parallel to the $Z$-axis, see 
\autoref{sec:efd_BO_p}.

\begin{figure}[h]
\begin{center}
\includegraphics[angle=0,width=0.95\columnwidth]{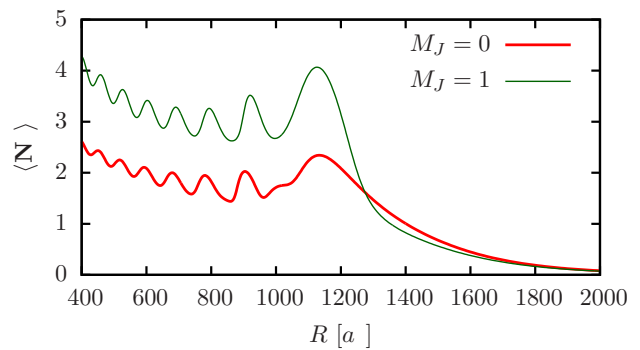}
\caption{For the KRb diatomic molecule within the  Rb-KRb \ry molecule the expectation value $\langle\mathbf{N}^2\rangle$
versus the separation $R$ between the Rydberg core and KRb is shown.}%
\label{fig:n2_data}
\end{center}
\end{figure}
For a rotating molecule in an electric field, the pendular states appear for strong fields when it becomes oriented
along the field direction; each pendular state being a coherent superposition
of field-free rotational states. 
For the diatomic KRb within the triatomic Rb-KRb, the electric field induced by the \ry core and electron should
be expected to lead to these pendular levels. The  strong hybridization of the angular motion of  KRb 
is  illustrated by  the expectation value  
$\langle\mathbf{N}^2\rangle$. For the lowest triatomic molecule states with $M_J=0$ and $1$ evolving from the Rb($n=24$, $l\ge 3$) degenerate manifold,  
$\langle\mathbf{N}^2\rangle$ is plotted as a function of  $R$ in \autoref{fig:n2_data}.
For both states, $\langle\mathbf{N}^2\rangle$ oscillates reaching values well above the field-free one $\langle\mathbf{N}^2\rangle=0$.
The fact that $\langle\mathbf{N}^2\rangle$ surpasses the field-free value of  the rotational state $N=1$, 
indicates that higher rotational excitations contribute to the field-dressed rotational dynamics of KRb. 
As $R$ is further increased beyond $1200~a_0$, the electric field strength is reduced and $\langle\mathbf{N}^2\rangle$ decreases 
slowly approaching zero for $R\gtrsim 2000~a_0$.  
This indicates that these states belong to the degenerate manifold formed by the
$n^2-9$ levels  of the \ry atom and KRb in its rotational ground state. For alkali metal atoms in general, there are three Rydberg levels with appreciable quantum defects, $s$, $p$, and $d$ levels, accounting for a total of $n^2-9$ hydrogenic levels.

\begin{figure}[h]
\begin{center}
\includegraphics[angle=0,width=0.9\columnwidth]{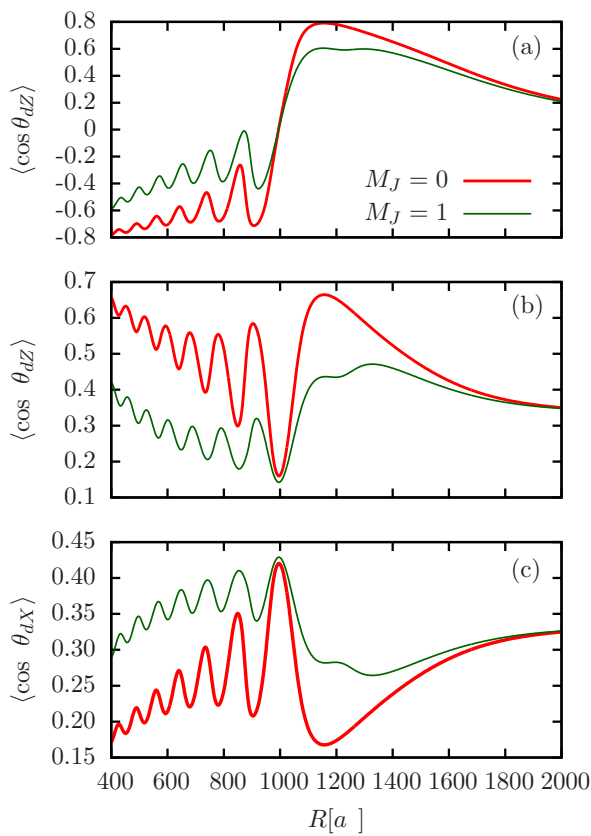}
\caption{For the KRb diatomic molecule within the Rb-KRb triatomic \ry molecule, we show the (a) orientation  
$\langle\cos\theta_{dZ}\rangle$, alignment along (b) the  $Z$-axis  $\langle\cos^2\theta_{dZ}\rangle$
and (c)  the $X$-axis  $\langle\cos^2\theta_{dX}\rangle$ versus the separation between the
Rydberg core and KRb. The results are presented for the triatomic molecule states with $M_J=0$ (thick line) and $M_J=1$ (thin line) evolving from
the Rb($n=24$, $l\ge 3$) degenerate  manifold.}%
\label{fig:orien_alig}
\end{center}
\end{figure}
The KRb molecule is strongly oriented and aligned due to the electric field induced by 
 the \ry core and electron. 
The orientation and alignment of the diatomic molecule along the  LFF Z-axis
in the lowest triatomic molecule states with $M_J=0$ and $1$ for the Rb manifold $n=24$ and $l\ge 3$,  
are illustrated by the expectation values 
$\langle\cos\theta_{dZ}\rangle$ 
and 
$\langle\cos^2\theta_{dZ}\rangle$ 
in  \autoref{fig:orien_alig}~(a) and  \autoref{fig:orien_alig}~(b), respectively.
Since the \ry electric field has also components along the $X$ and $Y$-axes, 
we have also estimated the alignment of KRb  along the LFF $X$-axis by means of  
$\langle\cos^2\theta_{dX}\rangle$ in   \autoref{fig:orien_alig}~(c).  
 Let us mention that   
 the alignments along the $Y$ and $X$ axes are identical, \ie, $\langle\cos^2\theta_{dX}\rangle=\langle\cos^2\theta_{dY}\rangle$, 
and that  the diatomic molecule does not gain any orientation along these axes,
 \ie,  $\langle\cos\theta_{dX}\rangle=\langle\cos\theta_{dY}\rangle=0$.
The  KRb within these two states of the triatomic molecule
is strongly oriented toward the \ry core for $R\lesssim 1000~a_0$ and  %done
$\langle\cos\theta_{dZ}\rangle$ oscillates as $R$ is varied.
As $R$ is increased, $\langle\cos\theta_{dZ}\rangle$ decreases and the dipole becomes oriented away from the  \ry core. %done
The maximal values of $\langle\cos\theta_{dZ}\rangle$ are $0.78$ and  $0.60$, for the $M_J=0$ and $1$, respectively,  % done
$\langle\cos\theta_{dZ}\rangle$
monotonically decreases for  $R\gtrsim 1200a_0$ approaching the field-free value %done
$\langle\cos\theta_{dZ}\rangle=0$. Note that even for $R\approx2000~a_0$
these states present a weak orientation with  $\langle\cos\theta_{dZ}\rangle\approx0.2$. %done
The alignment parameters, $\langle\cos^2\theta_{dZ}\rangle$ and $\langle\cos^2\theta_{dX}\rangle$
also show oscillatory behavior as  $R$ is varied. 
For the $M_J=0$ state, the diatomic molecule is more aligned along the $Z$ axis than along the $X$-axis,
\ie, $\langle\cos^2\theta_{dZ}\rangle>\langle\cos^2\theta_{dX}\rangle$.
Whereas, for  $M_J=1$ state, it holds $\langle\cos^2\theta_{dZ}\rangle<\langle\cos^2\theta_{dX}\rangle$.
For both states, at those values of $R$ where $\langle\cos^2\theta_{dZ}\rangle$ reaches a maximum, 
$\langle\cos^2\theta_{dX}\rangle$ reaches a minimum and viceversa. 
The smallest (largest) value of $\langle\cos^2\theta_{dZ}\rangle$ ($\langle\cos^2\theta_{dX}\rangle$) is obtained when 
$\langle\cos\theta_{dZ}\rangle\approx0$, \ie,
 KRb changes from being oriented  toward the Rb$^+$ core to away from it. % done 
The maxima and minima of $\langle\cos^2\theta_{dZ}\rangle$ and  $\langle\cos\theta_{dZ}\rangle$
are reached at very close values of $R$.

For the $M_J=1$ triatomic molecule state,
$\langle\cos\theta_{dZ}\rangle$ and  $\langle\cos^2\theta_{dZ}\rangle$ reach 
a shallow minimum and  $\langle\cos^2\theta_{dX}\rangle$  a maximum for $R\approx 1200~a_0$
which are due to a very broad avoided crossing  that this state suffers with the neighboring one. 
Note that in the behavior of $\langle\mathbf{N}^2\rangle$, this avoided crossing is not observed because both involved triatomic molecule states
have similar values of  $\langle\mathbf{N}^2\rangle$.

\section{The electric field-dressed Born-Oppenheimer potentials}
\label{sec:efd_BO_p}

In this section, we investigate the impact of an additional external electric field on the BOPs of the  Rb-KRb triatomic molecule.  
For the sake of simplicity, we assume that this external field is antiparallel to the LFF $Z$-axis,
\ie,  $\mathbf{F}_{ext}=-F_{ext}\hat{Z}$.  
For a tilted external electric field, the  azimuthal symmetry will be broken, and $M_J$ will no longer be a good quantum number, increasing substantially the overall complexity. We focus on the weak field regime  $F_{ext}\le 12$~V/cm.
For such weak fields, the impact on the KRb rotational motion is negligible
because  the shift due the field, $dF_{ext} \le 3.1$~MHz for  $F_{ext}\le 10$~V/cm,  is much smaller than $B=1.114$~GHz.
In contrast,  due to its large dipole moment, the \ry levels of rubidium are significantly affected.
This implies that the electric field induced by the \ry ion and electron in the diatomic molecule is also modified and as a consequence 
the level structure of the triatomic \ry molecule and the BOP change correspondingly.

\begin{figure}[h]
\begin{center}
\includegraphics[angle=0,width=0.95\columnwidth]{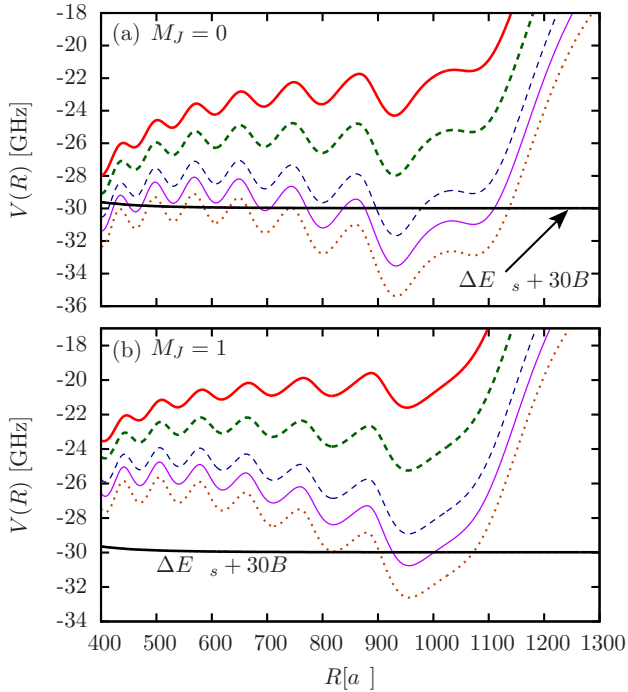}
\caption{Rb-KRb triatomic molecule in an external electric field. The lowest lying Born-Oppenheimer potentials are shown for
(a) $M_J=0$ and (b) $M_J=1$ evolving from the Rb($n=24$, $l\ge3$) \ry manifold 
as a function of the separation between Rb$^+$ and KRb. The electric field strengths are
$F_{ext}=0$~V/cm (thick solid), $4$~V/cm (thick dashed), $8$~V/cm (thin dashed), $10$~V/cm (thin  solid), 
and $12$~V/cm (dotted).}%
\label{fig:BOP_field}
\end{center}
\end{figure}
In \autoref{fig:BOP_field}(a) and (b), we present the lowest BOP with $M_J=0$ and $1$, respectively, evolving  from the 
Rb($n=24$, $l\ge 3$)  manifold with increasing electric field strength $F_{ext}$. 
As the field becomes stronger, the BOPs are lowered,  but the reduction in energy depends on the separation 
between the Rb atom and the KRb molecule, and the smaller $R$, the smaller  this reduction is. 
Thus,  the lowest BOP evolving  from the Rb($n=24$, $l\ge 3$) \ry manifold will at some point cross
the BOP from Rb($27s$) - KRb($N=5$) level, creating a sequence of avoided crossings
which are narrow due to the weak coupling between the involved levels. 
As  a consequence of the $R$-dependent effect of the external field, the widths of the outermost potential well is increased. This
is particularly observed in  the last two outer wells for the $M_J=0$ BOP, where for $F_{ext}\ne0$ a double-well structure becomes
visible. 

\section{Vibrational states}
 \label{sec:BO_p_vib}
  
\begin{figure}[t]
\begin{center}
\includegraphics[angle=0,width=0.95\columnwidth]{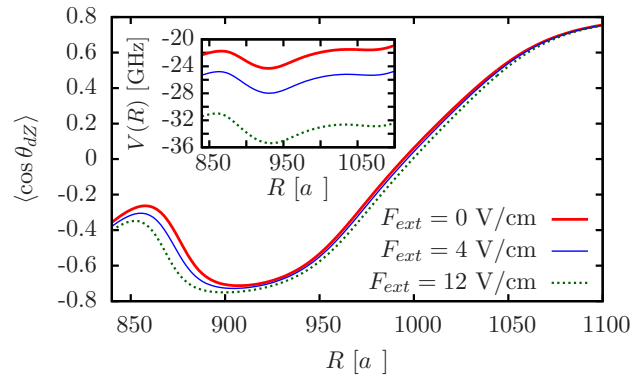}
\caption{Rb-KRb triatomic molecule in an external electric field for $F_{ext}=0$, $4$ and $12$~V/cm, 
the orientation $\langle\cos\theta_{dZ}\rangle$ of KRb within the lowest lying BOP with $M_J=0$ 
evolving from the degenerate \ry manifold $n=24$ and $l\ge3$ of Rb is shown. In the inset the energy of these BOP versus $R$
is provided.}%
\label{fig:BOP_MF_0_orientacion}
\end{center}
\end{figure}

In this section, we present the vibrational bound levels within the outer two minima 
of the BOP belonging to the lowest $M_J=0$ triatomic molecule state evolving from the Rb($n=24$, $l\ge3$) degenerate manifold, in  an external electric field of strength $F_{ext}=0$, $4$ and $12$~V/cm.
The perturber molecule KRb within the Rb-KRb triatomic system is oriented towards the Rb$^+$ core in the inner (L) well,
and becomes  oriented away from it in the last shallow well, see \autoref{fig:BOP_MF_0_orientacion}. 
By increasing $F_{ext}$, the orientation of KRb is only weakly affected, but the depth of the BOP is increased,
which implies an enhancement of the number of bound vibrational levels.
In \autoref{fig:BOP_MF_0_orientacion}, one can appreciate the shift between the minima 
in the BOP, and the maxima/minima in the orientation of the KRb diatomic molecule.
 
\begin{figure}[t]
\begin{center}
\includegraphics[angle=0,width=0.95\columnwidth]{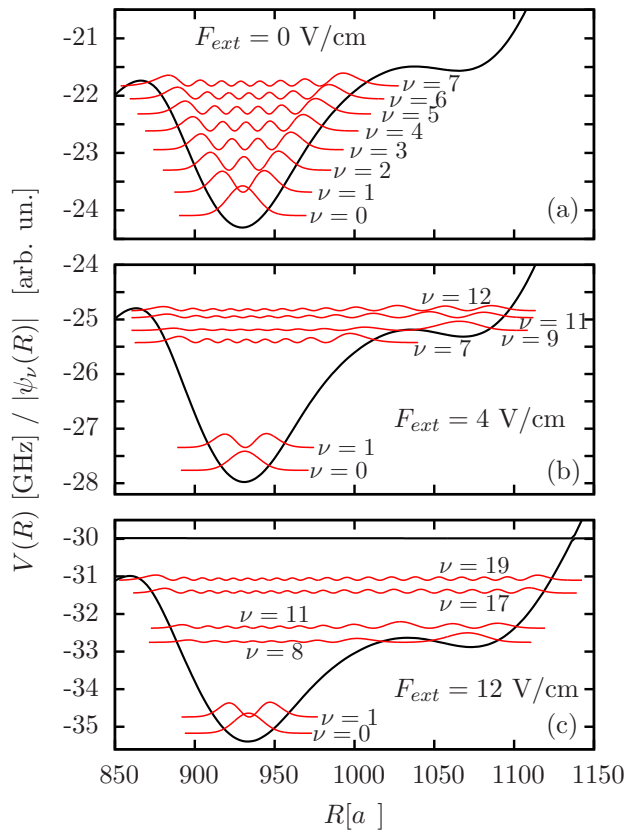}
\caption{Modulus  square of the wave functions $|\psi_\nu(R)|^2$ of several vibrational bound states 
within the last two  wells  of the energetically lowest BOP with $M_J=0$
evolving from the degenerate  Rb($n=24$, $l\ge3$) \ry manifold. 
The square of the wave  functions are plotted  in arbitrary units, and their positions are shifted to the energies of the corresponding 
vibrational state.}%
\label{fig:BOP_MF_0_vibrational_F0}
\end{center}
\end{figure}

\begin{figure}[t]
\begin{center}
\includegraphics[angle=0,width=0.95\columnwidth]{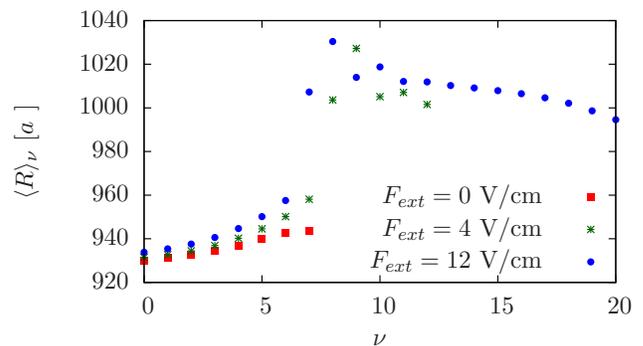}
\caption{For an external electric field $F_{ext}=0$, $4$ and $12$~V/cm, 
we show the expectation value $\langle R \rangle_\nu$ of the vibrational bound states 
within the last two wells  of the lowest BOP with $M_J=0$
evolving from the degenerate \ry manifold, Rb($n=24$, $l\ge3$).}%
\label{fig:BOP_MF_0_R}
\end{center}
\end{figure}

With no external field, the last (R) outer well potential is not deep enough to accommodate 
a bound state.  In the inner potential well (L), there exists eight bound vibrational levels.
In \autoref{fig:BOP_MF_0_vibrational_F0}~(a), we present the absolute square of the wave functions of these vibrational states. 
The expectation values $\langle R \rangle_\nu=\langle \psi_{\nu,0}|R| \psi_{\nu,0} \rangle$, 
with $ \psi_{\nu,0}= \psi_{\nu,0}(R)$ being the vibrational wave functions, are  presented 
versus the vibrational quantum number $\nu$ in  \autoref{fig:BOP_MF_0_R}. 
%These values of  $\langle R \rangle_\nu$ indicate that the vibrational levels are located in the second-to-last minimum.
These vibrational bands, with an energy spacing of a few  hundreds MHz, \ie, $|E_{\nu=1}-E_{\nu=0}|=408$~MHz
and  $|E_{\nu=7}-E_{\nu=6}|=183$~MHz, 
have  superimposed rotational structure determined 
by the rotational constant $B_\nu=\langle R^{-2} \rangle_\nu/2\mu$, of the order of  a few tenths kHz, \ie,
$B_0=40.2$~kHz and $B_7=37.5$~kHz. 

For $F_{ext}=4$~V/cm, there are $13$ vibrational levels bound in the two outermost potential wells.
Due to the presence of the external field, the last well becomes deeper and can accommodate vibrational levels, and accordingly
we encounter
in \autoref{fig:BOP_MF_0_vibrational_F0}~(b) a double-well structure.
For instance, the wave functions of the states with vibrational quantum number  $v=8,\dots,  12$ extend over both wells
with $\langle R \rangle_\nu>1000~a_0$,  see \autoref{fig:BOP_MF_0_R}. 
For $R\approx 1000~a_0$, the orientation of KRb 
is close to zero, since its dipole is changing from being oriented toward  the Rb$^+$ core %done 
to away from it. Thus,
only for the $\nu=9$ state, KRb is significantly oriented toward the \ry core. 
A similar energy spacing is found between neighboring vibrational levels as in the $F_{ext}=0$~V/cm case, except
for the $\nu=8$ and $9$ states, which are $36$~MHz apart. 

By further increasing the field to  $F_{ext}=12$~V/cm, the last two potential wells possess $20$ vibrational 
levels. The vibrational wave function of the $\nu=8$  state is partially contained in the outermost (R) well, cf. 
\autoref{fig:BOP_MF_0_vibrational_F0}~(c), having $\langle R \rangle_\nu\approx1030~a_0$,
and  at  this distance  the polar molecule is oriented away from the rubidium core.
Highly excited vibrational states have  $\langle R \rangle_\nu\approx1000~a_0$,
and for similar separation distances,   the KRb 
molecule does not present any orientation, see \autoref{fig:BOP_MF_0_orientacion}.
The smallest vibrational spacing of $12.5$~MHz is found between the $v=7$ and $8$ levels.

 Due to the non-adiabatic interactions, the 
 electronic states are hybridized allowing for intra-electronic dipole transitions.
 For instance, for  $F_{ext}=12$~V/cm ($4$~V/cm)
the vibrational levels $\nu=0$ and $\nu=8$ ($\nu=9$) could be coherently coupled in a Raman process via  a
highly excited vibrational state within this BOP as has been suggested in Ref.~\cite{rittenhouse10}. 
In such a way, the KRb orientation becomes entangled. 

%%%%%%%%%%%%%%%%%%%%%%%%%%%%%%%%%%%%%%
\section{Conclusions}
\label{sec:con}

We have investigated ultralong-range triatomic \ry molecules formed by a \ry rubidium atom and the KRb diatomic rotational molecule in the presence of electric fields. The field effect arises from both the "internal" \ry core and electrons and from an externally applied field. This species exhibits novel binding properties due to the attractive interaction
of the \ry electron and the ground state KRb diatomic molecule, which can be controlled by an external field.
Previous studies have modeled the diatomic molecule in a  simple two level paradigm. We have been systematically
extending this approach by taking into account the rotational motion of the diatomic molecule which couples, due to its dipole moment,
to the electric field supplied by the \ry ionic core and the Rydberg electron or an external source.
The rotational motion of the KRb diatomic molecule is described within the rigid rotor approximation, which allows for a proper description of the
hybridization of the rotational motion of KRb due to the electric fields.
The Born-Oppenheimer potential curves for $M_J=0,1$ for the \ry triatomic molecule as a 
function of the separation between the Rb$^+$ core and the KRb diatomic molecule have been derived and analyzed. 
We have performed detailed analysis of the hybridization of the angular motion, orientation and alignment of KRb within the Rb-KRb triatomic molecule. These results demonstrate that excited rotational states are involved in the field-dressed dynamics, and, therefore, the rotational
degrees of freedom are needed for a proper description of the \ry triatomic molecule.

In an additional external electric field, the level structure of the \ry triatomic molecule is severely modified.  The BOPs evolving from the Rb($n=24$, $l\ge 3$) \ry manifold are strongly lowered, due to electronic hybridization, as $F_{ext}$ increases, whereas those levels evolving from the non-degenerate Rb($27$) state are weakly affected. The field-dressed potentials strengthen the bound state character of the
rovibrational molecular states and lead for certain configurations to narrow avoided crossings.
In the presence of an external field, a Raman scheme among different vibrational levels within the lowest electronic state evolving from the 
Rb($n=24$, $l \ge 3$) Rydberg manifold  could be
used to prepare entangled states of KRb levels with different orientations.

In an ultracold quantum gas which forms the KRb, there are typically as many potassium atoms as rubidium atoms.
Hence, there are two options to form the triatomic \ry molecular states, \ie K$^*$-KRb and Rb$^*$-KRb
triatomic molecules. The main difference between the two highly excited triatomic molecules are the different low angular
momentum states with appreciable quantum defects and energy levels of the \ry spectra of K and Rb, which provide a variable binding of the corresponding ground state
heteroncuclear diatomic molecule KRb. The two types of triatomic molecules do share the atomic (K or Rb) degenerate manifold such as the $n=24$ and $l\ge 3$, as employed here. Thus, the Born-Oppenheimer potentials emerging from the degenerate manifold are identical
for both triatomic molecules since they only depend on the properties of the polar ground state diatomic molecule and the atomic \ry degenerate manifold under consideration. 
Their interaction (avoided crossings) with the quantum defect Born-Oppenheimer potentials will, of course,
move in energy and nuclear configuration space from one to the other species. Similar arguments hold for
any other combination of \ry atoms and polar diatomic molecules possessing an electric dipole moment below the critical value 
$d_{cr}=0.639315~$a.u \cite{fermi47,turner77,clark79,fabrikant04} above which 
the \ry electron would bind to the dipolar diatomic molecule.

An even more complete description of the triatomic \ry molecule would take into account the vibrational motion of the diatomic molecule. 
However, the energy scales involved into the vibrations are much larger than those of the rotation and therefore
little impact of the electric field of either internal or external origin can be expected. Also, the coupling of
the rotation to the vibration of the diatomic molecule should be negligible for sufficiently low-lying rovibrational states.
The corresponding potential energy surfaces then become four-dimensional with three vibrational degrees of freedom
from the field-free situation and the additional rotational degree of freedom which turns into a vibration for sufficiently
strong fields.

\begin{acknowledgments}

R.G.F. gratefully acknowledges a Mildred Dresselhaus award from the excellence
cluster "The Hamburg Center for Ultrafast Imaging Ð Structure, Dynamics and Control of Matter at the
Atomic Scale" of the Deutsche Forschungsgemeinschaft and financial support by the Spanish Ministry of Science FIS2011-24540 (MICINN), grants
P11-FQM-7276 and FQM-4643 (Junta de Andaluc\'{\i}a), and by the Andalusian research group FQM-207.
We also acknowledge financial support by the Initial Training Network COHERENCE of the European Union FP7 framework. H.R.S. and P.S. acknowledge ITAMP at the Harvard-Smithsonian Center for Astrophysics for support. 
\end{acknowledgments}

\end{document}